\documentclass{article}
\usepackage{stywhispers}
\usepackage{dsfont}
\usepackage{amsfonts}
\usepackage{amssymb}
\usepackage{amsmath,bm}
\usepackage{cases}
\usepackage{graphicx,cite,epsfig}
\usepackage{afterpage}
\usepackage{multirow}
\usepackage{algorithm}
\usepackage{algorithmic}
\usepackage{array}
\usepackage{subfigure}
\usepackage{amsmath,boxedminipage}
\usepackage{CJK}
\usepackage{hyperref}
\usepackage{enumerate}
\usepackage{setspace}
\usepackage{booktabs}

\usepackage{indentfirst}

\newcommand{\jbhao}{\fontsize{9.9pt}{\baselineskip}\selectfont}

\title{An Approximate Message Passing Approach For Compressive Hyperspectral Imaging using a Simultaneous Low-Rank and Joint-Sparsity Prior}%

\name{Yangqing Li$^{\star}$, Saurabh Prasad$^{\dagger}$, Wei Chen$^{~\!\prime}$, Changchuan Yin$^{\star}$, and Zhu Han$^{\dagger}$ \vspace{-12pt} }

\address{\jbhao $^{\star}$ Beijing Laboratory of Advanced Information Network, Beijing University of Posts and Telecommunications, Beijing, China \\
    \jbhao $^{\dagger}$ Electrical and Computer Engineering Department, University of Houston, USA\\
    \jbhao $^{~\!\prime}$ State Key Laboratory of Rail Traffic Control and Safety, Beijing Jiaotong University, Beijing, China \thanks{
    The work of S. Prasad was funded in part by the NASA New Investigator (Early Career) Award, grant \# NNX14AI47G.
    The work of Z. Han was partially supported by the U.S. National Science Foundation under grants ECCS-1547201, CCF-1456921, CNS-1443917, ECCS-1405121, CMMI-1434789, and NSFC 61428101.}\vspace{-12pt}}

\begin{document}

\maketitle

\vspace{-6pt}
\begin{abstract}
This paper considers a compressive sensing (CS) approach for hyperspectral data acquisition, which results in a practical compression ratio substantially higher than the state-of-the-art. Applying simultaneous low-rank and joint-sparse (L\&S) model to the hyperspectral data, we propose a novel algorithm to joint reconstruction of hyperspectral data based on loopy belief propagation that enables the exploitation of both structured sparsity and amplitude correlations in the data. Experimental results with real hyperspectral datasets demonstrate that the proposed algorithm outperforms the state-of-the-art CS-based solutions with substantial reductions in reconstruction error.
\vspace{-2pt} \end{abstract}

\begin{keywords}
Compressive hyperspectral imaging, low-rank and joint-sparse, compressive sensing, approximate message passing
\vspace{-14pt} \end{keywords}

\section{Introduction}
\vspace{-8pt}
\label{section1}
Unlike traditional imaging systems, hyperspectral imaging (HSI) sensors \cite{hyper09},\cite{hyper13} acquire a scene with several millions of pixels in up to hundreds of contiguous wavelengths. Such high resolution spatio-spectral hyperspectral data, i.e., three-dimensional (3D) datacube organized in the spatial and spectral domain, has an extremely large data size and enormous redundancy, which makes compressive sensing (CS) \cite{CS06} a promising solution for hyperspectral data acquisition.

To date, most existing designs for CS-based hyperspectral imagers can be grouped into frame-based acquisition in the spatial direction \cite{hypersensing08,hypersensing12,L&Shypersensing12,hypersensing15} and pixel-based acquisition in the spectral direction \cite{struchyper14,push&whisk14,whisk09}. While a lot of reconstruction approaches for these two acquisition schemes have been proposed, most existing algorithms can only take advantage of the spatial and spectral information of hyperspectral data from the aspect of sparsity (or joint-sparsity). Because the foundation of these algorithms is built on conventional CS, which reconstructs the signals by solving a convex programming and proceeds without exploiting additional information (aside from sparsity or compressibility) \cite{CS06}. For hyperspectral data, the spatial and spectral correlations, which not only reflect in the correlation between the sparse structure of the data (i.e., structured sparsity), but also in the correlation between the amplitudes of the data, can be used to provide helpful prior information in the reconstruction processed and assist on increasing the compression ratios.

In this paper, the structured sparsity and the amplitude correlations are considered jointly by assuming that spatially and spectrally correlated data satisfies simultaneous low-rank and joint-sparse (L\&S) structure. Using a structured L\&S factorization, we propose an iterative approximate message passing (AMP) algorithm \cite{MP09},\cite{GAMP11}, in order to enable joint reconstruction of the data with the practical compression ratio that is substantially higher than the state-of-the-art. Specifically, in section \ref{section2}, we introduce the structured factorization representation of the L\&S model. In section \ref{section3}, we propose a novel AMP-based approach, called L\&S-approximate message passing (L\&S-AMP), that decouples the global inference problem into two sub-problems. One sub-problem considers the linear inverse problem of recovering the signal matrix from its compressed measurements. Another sub-problem exploits the L\&S structure of the signal matrix. Then a recently proposed ``turbo AMP" framework \cite{TurboAMP10} is used to enable messages to pass between these two phases efficiently. Section \ref{section4} presents simulation results with real hyperspectral data that support the potential of the approach to considerably reduce the reconstruction error. In section \ref{section5}, we conclude the paper.

\vspace{-8pt}
\section{Problem Setup}
\vspace{-6pt}
\label{section2}
In this section, we first present the problem for compressive hyperspectral imaging. Then, we propose a structured L\&S factorization model for the signal matrix, which will be later exploited to acquire any HSI with very few measurements, via a novel joint reconstruction approach.

\vspace{-8pt}
\subsection{Data Acquisition Model}
\vspace{-6pt}
Owing to the inherent 3D structure present in the hyperspectral datacube and the two-dimensional nature of optical sensing hardware, CS-based hyperspectral imagers generally capture a group of linear measurements across either the 2D spatial extent of the scene for a spectral band or the spectral extent for a spatial (pixel) location at a time, i.e., $\textbf{f}_t\in\mathbb{R}{^{N}}, t\in[1,...,T]$. Then, the compressed measurements $\textbf{y}_1, \textbf{y}_2,..., \textbf{y}_T$ are sent to a fusion station that will recover the original 3D datacube by utilizing a CS reconstruction algorithm.

By taking the highly correlated hyperspectral data vectors $\textbf{f}_1,\textbf{f}_2,\cdots,\textbf{f}_T$ to admit a sparse representation in some orthonormal basis, e.g., DCT basis or wavelet basis, we have,
\vspace{-7pt}
\begin{equation}\label{sparse}
  \textbf{f}_t=\bm{\Psi}\textbf{x}_t,~~~t=1,\ldots,T,\
\vspace{-4pt}
\end{equation} where sparse signal vectors $\textbf{x}_t\in\mathbb{R}^{N},\forall t$. $\bm{\Psi} \in\mathbb{R}{^{N \times N}}$ is a certain orthonormal basis matrix. Then we obtain the typical CS formulation as follows
\vspace{-4pt}
\begin{equation}\label{cs}
    \textbf{y}_t=\textbf{A}_t\textbf{x}_t+\textbf{w}_t=\textbf{z}_t+\textbf{w}_t,~~~t=1,\ldots,T,\
\vspace{-4pt}
\end{equation}where $\textbf{A}_t = \bm{\Phi}_t\bm{\Psi} = [{a_{mn}}]\in\mathbb{R}^{M_t\times N}$, $\textbf{z}_t\in\mathbb{R}^{M_t}$ is the measurement output vector, and $\textbf{w}_t$ is an additive noise vector with unknown variance $\tau _t^\omega$.

\vspace{-8pt}
\subsection{Structured L\&S factorization for signal matrix}
\vspace{-6pt}
\label{section2.3}
As mentioned in the introduction, while the original hyperspectral data can be reconstructed by using conventional CS recovery algorithms, it is possible to achieve a much better recovery performance by applying the L\&S model to further exploit the structural dependencies between the values and locations of the coefficients of the sparse signal vectors $\textbf{x}_t,\forall t$. The main reason that we consider $\textbf{X}$ as a L\&S matrix is two-fold. First, images from different spectral bands enjoy similar natural image statistics, and hence can be joint-sparse in a wavelet/DCT basis \cite{L&Shypersensing12}; second, a limited number of unique materials in a scenes implies that spectral signatures across pixels can be stacked to form a matrix that is often low-rank \cite{HU_AMP15}.

To precisely achieve the benefits of the L\&S model and reconstruct the original hyperspectral data from a Bayesian point of view, here we propose an accurate probabilistic model by performing a structured L\&S factorization for $\textbf{X}$ as
\vspace{-4pt}
\begin{equation}\label{SignalModel}
  \textbf{X}\triangleq\textbf{S}\bm{\Theta}\triangleq\textbf{SHL}\triangleq\textbf{GL},
\vspace{-4pt}
\end{equation}where the diagonal matrix $\textbf{S}={\rm{diag}}(s_1,s_2,\cdots,s_N)$ is the sparsity pattern matrix of the signals with the support indicates ${s_n} \in \left\{ {0,1} \right\},~\forall n$. We refer to $K\!=\!\sum\nolimits_{n = 1}^N {{s_n}} \!\ll\! N$ as the sparsity level of $\textbf{X}$. $\textbf{H}= [{h_{nr}}]\in\mathbb{R}^{N\times R}$ and $\textbf{L}= [{l_{rt}}]\in\mathbb{R}^{R\times T}$ are obtained from the low-rank matrix factorization of $\bm{\Theta}\in\mathbb{R}^{N\times T}$, which is the amplitude matrix of $\textbf{X}$. For a joint-sparse matrix ${\textbf{G}}$ and an arbitrary matrix $\textbf{L}$, this factorization implies that $\textbf{X}$ is a simultaneous low-rank ($R\leq K\ll N$) and joint-sparse matrix with rank $R \le {\rm{min}}(K,T)$, where all sparse signal vectors $\textbf{x}_1,~\textbf{x}_2,\!~\cdots~\!,\textbf{x}_T$ share a common support with sparsity level $K$.

Assuming independent entries for $\textbf{S}$, $\textbf{H}$, and $\textbf{L}$, the separable probability density functions (PDFs) of $\textbf{G}$ and $\textbf{L}$ become
\vspace{-4pt}
\begin{equation}
p(\textbf{G})\!=\!\prod\nolimits_{n,r} p(g_{nr})\!=\!\prod\nolimits_n\!\!\left(\!p(s_n)\!\prod\nolimits_r\!\mathcal{N}(h_{nr};\hat g_0,\nu _0^g)\!\right)\!,
\vspace{-2pt}
\end{equation}
\begin{equation}
\hspace{-69pt}
p(\textbf{L}) = \prod\nolimits_{r,t} p({l_{rt}})  = \prod\nolimits_{r,t} {\mathcal{N}({l_{rt}};0,1)} ,
\vspace{-4pt}
\end{equation}
\noindent
where both $\{h_{nr}\}_{\forall n,r}$ and $\{l_{rt}\}_{\forall r,t}$ are assumed to be i.i.d. Gaussian with unknown mean ${\hat g_0}$ and variance $\nu _0^g$. In particular, we assume $\{l_{rt}\}_{\forall r,t}$ follow i.i.d. Gaussian distribution with zero mean and unit variance, i.e., $\mathcal{N}(0,1)$, to avoid ambiguity and the unnecessary model parameters update. As $\{s_n\}_{\forall n}$ are treated as i.i.d. Bernoulli random variables with $\Pr ({s_n} = 1) \buildrel \Delta \over = \lambda,~\forall n$, the sparse coefficients, $\{g_{nr}\}_{\forall n,r}$, become i.i.d. Bernoulli-Gaussian (BG), i.e., the marginal PDF
\vspace{-6pt}
\begin{equation}
p({g_{nr}}) = (1 - \lambda )\delta ({g_{nr}}) + \lambda \mathcal{N}({g_{nr}};{\hat g_0},\nu _0^g),~\forall n,r,~
\vspace{-2pt}
\end{equation}
\noindent
where $\delta(\cdot)$ is the Dirac delta function. Furthermore, due to the assumption of adaptive Gaussian noise in (\ref{cs}), the likelihood function of $\textbf{Z} = \{ \textbf{z}_t\}_{\forall t}$ is known and separable, i.e.,
\vspace{-4pt}
\begin{equation}
p(\textbf{Y}\left| \textbf{Z} \right.)\!=\!\prod\limits_{t = 1}^T {\prod\limits_{m = 1}^{M_t}\!{p\left( {{y_{mt}}\left| {{z_{mt}}} \right.} \right)} }\!=\!\prod\limits_{t = 1}^T {\prod\limits_{m = 1}^{M_t}\!{{\cal N}\left( {{y_{mt}};{z_{mt}},\tau _t^\omega } \right)} },
\vspace{-4pt}
\end{equation}where the measurement $\textbf{Y} = \{ \textbf{y}_t\}_{\forall t}$.

\begin{figure}[!t]
\setlength{\abovecaptionskip}{-1.5pt}
  \centering
  \includegraphics[width=7cm]{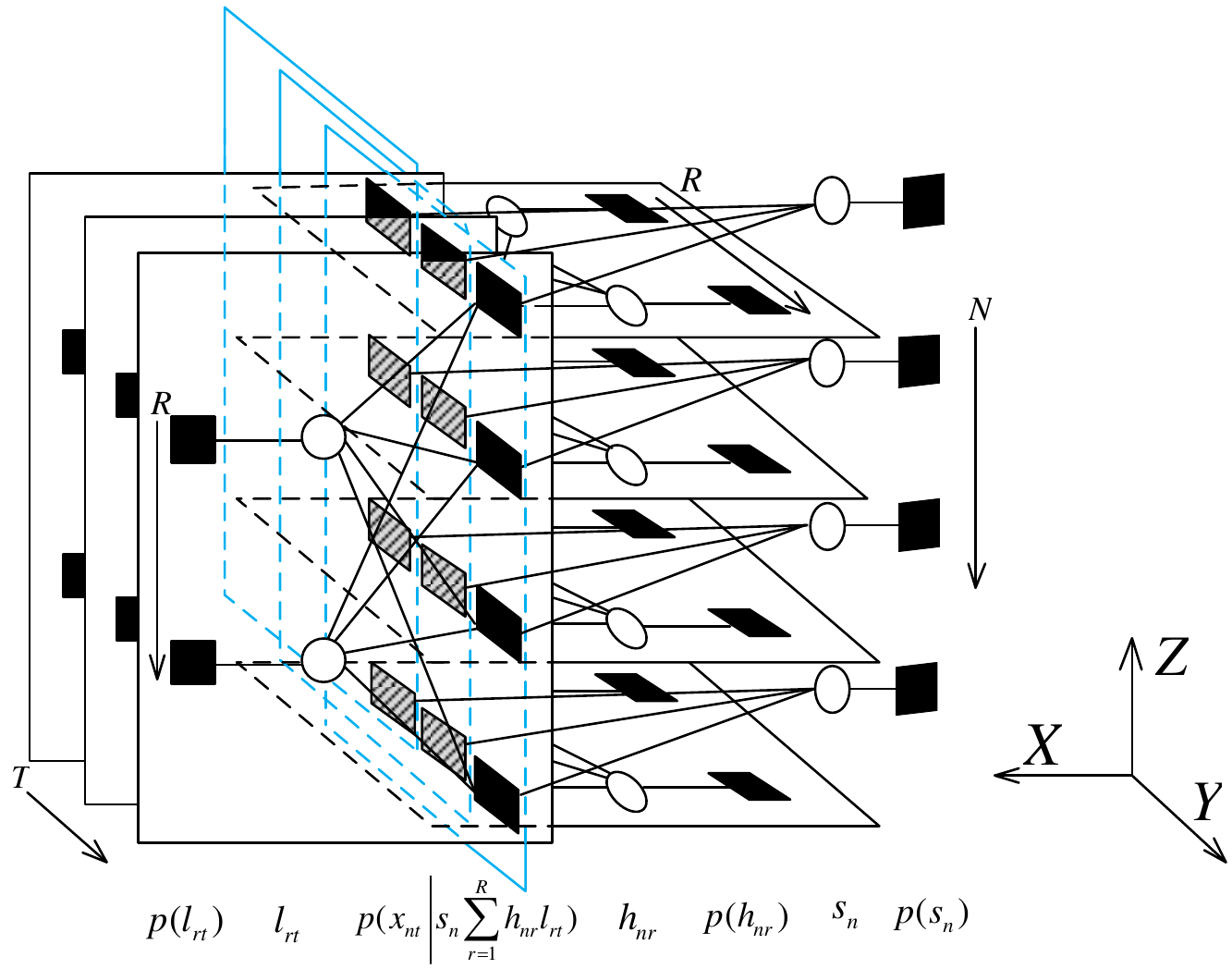}\vspace{-12pt}\\
  \caption{The factor graph for the L\&S structure model with $N = 4$, $T = 3$, $R = 2$, and $M = 3$.}\label{Fig_L&SAMP}
\vspace{-4pt}
\end{figure}

\vspace{-8pt}
\section{The L\&S-AMP Algorithm}
\begin{figure}[!t]
\setlength{\abovecaptionskip}{-1.5pt}
  \centering
  \includegraphics[width=6.3cm]{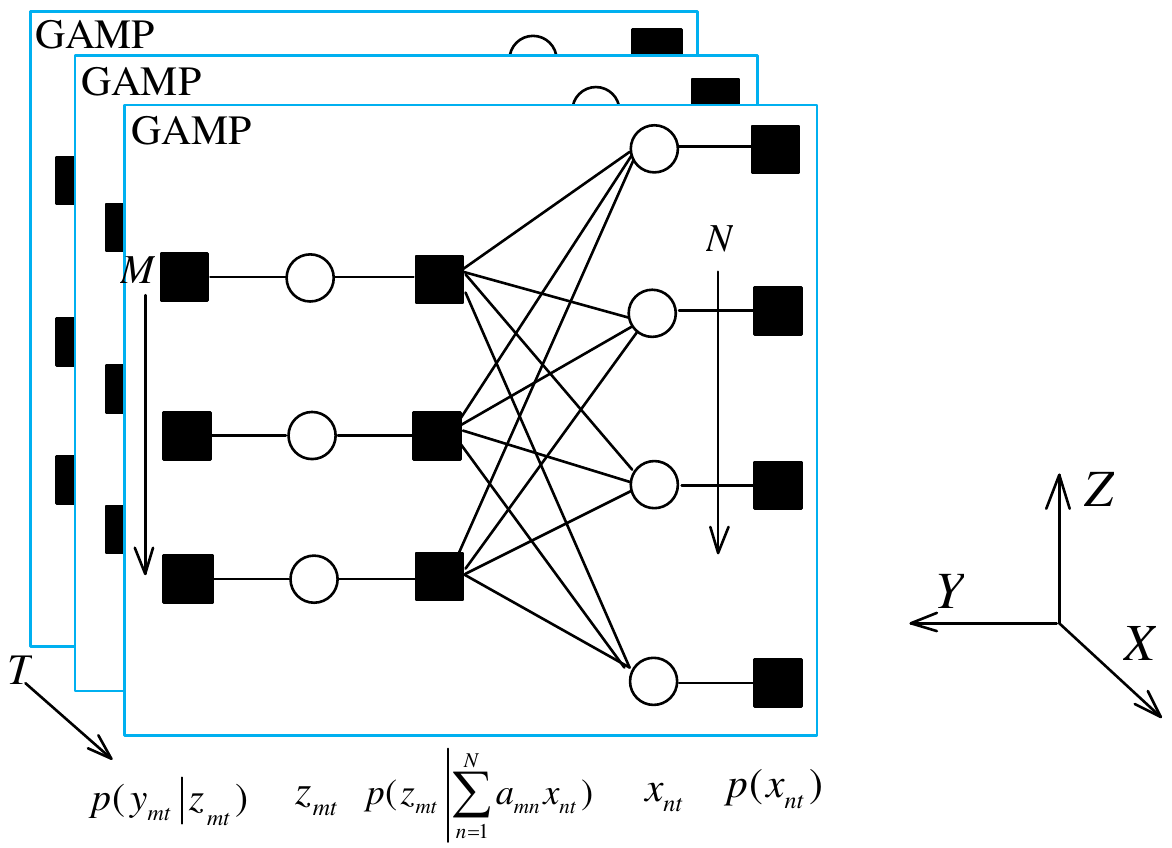}\vspace{-10pt}\\
  \caption{The factor subgraphs for the M-GAMP phase with $N = 4$, $T = 3$, $R = 2$, and $M = 3$.}\label{Fig_MGAMP}
\vspace{-10pt}
\end{figure}

\vspace{-6pt}
\label{section3}
With the problem formulation in (\ref{cs}) and (\ref{SignalModel}), our proposed method is to maximize the posterior joint distribution, i.e.,
\vspace{-8pt}
\begin{equation}\nonumber
\hspace{-20pt}
p(\textbf{S},\textbf{H},\textbf{L}\left| \textbf{Y} \right.{\rm{)}}=p(\textbf{Y}\left| \textbf{S},\textbf{H},\textbf{L}\!\!\right.{\rm{)}}p({\textbf{S},\textbf{H},\textbf{L}}){{} \mathord{\left/{{{p\textbf{(Y)}}}} \right.
\kern-\nulldelimiterspace}}
\end{equation}
\vspace{-18pt}
\begin{equation}\nonumber
\hspace{-95pt}
\propto\!\!~\prod\nolimits_t {p({\textbf{y}_t}\left| {{\textbf{A}_t}{\textbf{x}_t}} \right.)} p(\textbf{S})p(\textbf{H})p(\textbf{L})
\end{equation}
\vspace{-10pt}
\begin{equation}\nonumber
\hspace{-15pt}
=\prod\limits_{t = 1}^T{\prod\limits_{m = 1}^{M_t} {p\!\left(\!{{y_{mt}}\!\left| {\sum\limits_{n = 1}^N {\!{a_{mn}}{s_n}\!\!\sum\limits_{r = 1}^R {{h_{nr}}{l_{rt}}} } } \!\!\right.} \right)} }\!\times\!\prod\limits_{n = 1}^N {p({s_n})}
\end{equation}
\vspace{-6pt}
\begin{equation}\label{PJD}
\hspace{-66pt}
\times \prod\limits_{n = 1}^N {\prod\limits_{r = 1}^R {p({h_{nr}})} }  \times \prod\limits_{r = 1}^R {\prod\limits_{t = 1}^T {p({l_{rt}})} },
\end{equation}
\noindent
where $\varpropto$ denotes equality up to a normalizing constant scale factor. This posterior distribution can be represented with a factor graph shown in Fig. \ref{Fig_L&SAMP}, where circles denote random variables and squares denote posterior factors based on belief propagation \cite{FG01}. Each factor node represents the conditional probability distribution between all variable nodes it connected. $T$ vertical planes (parallel to $Y$ and $Z$ axes) exploit the linear measurement structure $\textbf{z}_t = \textbf{A}_t\textbf{x}_t, \forall t$ (detailed in Fig. \ref{Fig_MGAMP}), while the remaining part of Fig. \ref{Fig_L&SAMP} further exploits the L\&S structure $\textbf{X} = \textbf{SHL}$.

To bypass the intractable inference problem of marginalizing (\ref{PJD}), we propose to solve an alternative problem that consists of two sub-problems that mainly require local information to complete their tasks \cite{TurboAMP10}. Correspondingly, our proposed algorithm is divided into two phases: I) the \textit{multiple generalized approximate message passing} (M-GAMP) phase; II) the \textit{low-rankness and sparsity pattern decoding} (L\&SPD) phase. Owing to this, an efficient ``turbo AMP" iterative framework \cite{TurboAMP10} is used, that iteratively updates one of the phases' beliefs, and passes the beliefs to another phase, and vice versa, repeating until both phases converge.

\vspace{-8pt}
\subsection{M-GAMP Phase}
\vspace{-6pt}
\label{subsection3_1}
In each frame $t = 1, . . . , T$ of the M-GAMP phase, we apply the \textit{generalized approximate message passing} (GAMP) approach \cite{GAMP11} in parallel for the linear inference problem: estimate the vector $\textbf{x}_t$ from the observation $\textbf{y}_t$, as shown in Fig. \ref{Fig_MGAMP}. Specifically, the GAMP computes the approximated posteriors on ${\{ {x_{nt}}\} _{\forall n}}$ as \cite{MP09}, \cite{GAMP11}
\vspace{-6pt}
\begin{equation}\nonumber
\hspace{-60pt}
{\Delta _{{x_{nt}}}} = p({x_{nt}})\prod\nolimits_m {{\Delta _{{z_{mt}} \to {x_{nt}}}}}
\end{equation}
\vspace{-14pt}
\begin{equation}\label{Xpost}
\hspace{16pt}
\propto p({x_{nt}}|\sum\nolimits_r  {g_{nr}}{l_{rt}})\mathcal{N}({x_{nt}};{{\hat u}_{nt}},\nu _{nt}^u),
\end{equation}where ${\Delta _{A \to B}}$ denotes a message passed from node $A$ to the adjacent node $p(B)$ in the factor graph. The parameters ${{\hat u}_{nt}}$, and $\nu _{nt}^u$ are obtained after the GAMP iteration converges. For the prior distribution of ${x_{nt}}$, i.e., $p({x_{nt}}|\sum\nolimits_r {{g_{nr}}{l_{rt}}})$ used in (\ref{Xpost}), we can assume BG prior PDF
\vspace{-4pt}
\begin{equation}\label{Xmarginal}
{p({x_{nt}}|\!\sum\nolimits_r {{g_{nr}}{l_{rt}}})} \!=\! (1\!-\!\lambda )\delta (x_{nt})\!+\!\lambda \mathcal{N}({x_{nt}};{\hat q_{nt}},\nu_{nt}^q),
\vspace{-4pt}
\end{equation} where $\hat q_{nt}$ and $\nu_{nt}^q$ are the active-coefficient mean and the active-coefficient variance, respectively, of the variable ${x_{nt}}$. It is worth mentioning that the prior parameters $\{\hat q_{nt}\}_{\forall n,t}$ and $\{\nu_{nt}^q\}_{\forall n,t}$, are only initialized to agnostic values at the beginning of the L\&S-AMP algorithm (e.g., $\hat q_{nt} = 0, \nu_{nt}^q = 1, \forall n,t$), then iteratively updated according to the message passed from the L\&SPD phase. This process will be detailed in next subsection. Then the minimum-mean-squared error (MMSE) estimation of ${\{ {x_{nt}}\}_{\forall n,t}}$ is facilitated by the following prior-dependent integrals
\vspace{-6pt}
\begin{equation}\label{Xestimate}
\hspace{-50pt}
{\hat x_{nt}}(j) = \int {x_{nt}}{\Delta _{{x_{nt}}}}{\rm{d}}{x_{nt}}.
\vspace{-8pt}
\end{equation}

\subsection{L\&SPD Phase}
\vspace{-6pt}
\label{subsection3_2}
In the L\&SPD phase, to exploit the L\&S structure, we employ the recently proposed \textit{bilinear generalized approximate message passing} (BiG-AMP) approach \cite{BiGAMP14} to a variant of the PCA problem: estimate the matrices $\textbf{G}$ and $\textbf{X}$ from an observation $\textbf{\^{X}} = [{\hat x_{nt}}]\in\mathbb{R}^{N\times T}$ which is the posterior estimation of their product $\textbf{X}= \textbf{GL}$ obtained form the M-GAMP phase in (\ref{Xmarginal}). In particular, the BiG-AMP \cite{BiGAMP14} obtains the approximately Gaussian posterior messages ${\{ {\Delta _{x'_{nt}}}\} _{\forall n,t}}$ as
\vspace{-8pt}
\begin{equation}\nonumber
{\Delta _{{{x}'_{nt}}}}\!\!= p({{x}'_{nt}})\!\!\int_{{{\{ {g_{nr}}, {l_{rt}}\} }_{\forall r}}}\!\!\!{\left[ {\prod\nolimits_r {{\Delta _{{{x}'_{nt}} \leftarrow {g_{nr}}}}} } \right.}\!\!\times \!\!\left. {\prod\nolimits_r {{\Delta _{{{x}'_{nt}} \leftarrow {l_{rt}}}}} } \right]
\end{equation}
\vspace{-6pt}
\begin{equation}\label{Xpost1}
\hspace{-78pt}
\propto p({{x}'_{nt}}|\textbf{y}_t)\mathcal{N}({{x}'_{nt}};{{\hat q}_{nt}},\nu _{nt}^q),
\end{equation}
where the parameters ${{\hat q}_{nt}}$ and $\nu _{nt}^q$ are obtained after the BiG-AMP iteration converges. The prior distribution of ${x'_{nt}}$, i.e., $p({{x}'_{nt}}|\textbf{y}_t)$ used in (\ref{Xpost1}), comes from the posterior message of ${{x}_{nt}}$ given the observation $\textbf{y}_t$ in the M-GAMP phase. The prior distribution of $p({x_{nt}}|\sum\nolimits_r {{g_{nr}}{l_{rt}}})$ used in (\ref{Xpost}) comes from the posterior message of ${x_{nt}}$ given the matrix factorization ${x_{nt}}=\sum\nolimits_r {{g_{nr}}{l_{rt}}}$ in the L\&SPD phase.

To enable effective implementation of ``turbo AMP" iteration, given the construction of the factor graph in Fig. \ref{Fig_L&SAMP}, the sum-product algorithm (SPA) \cite{FG01} implies that,
\vspace{-4pt}
\begin{equation}\label{Xprior_SPA}
\hspace{-14pt}
p({{x}'_{nt}}|\textbf{y}_t) \propto \prod\nolimits_m {{\Delta _{{z_{mt}} \to {x_{nt}}}}({x_{nt}})},
\end{equation}
\vspace{-10pt}
\begin{equation}\nonumber
p({x_{nt}}|\sum\limits_r {{g_{nr}}{l_{rt}}} ) \propto \int_{{{\{ {g_{nr}}, {l_{rt}}\} }_{\forall r}}} {\left[ {\prod\nolimits_r {{\Delta _{{{x}'_{nt}} \leftarrow {g_{nr}}}}({g_{nr}})} } \right.}
\end{equation}
\vspace{-8pt}
\begin{equation}\label{Xprior_SPA1}
~~~~~~\times \left. {\prod\nolimits_r {{\Delta _{{{x}'_{nt}} \leftarrow {l_{rt}}}}({l_{rt}})} } \right].
\vspace{-4pt}
\end{equation}Comparing (\ref{Xprior_SPA}) and (\ref{Xprior_SPA1}) with (\ref{Xpost}) and (\ref{Xpost1}), respectively, we have
\vspace{-2pt}
\begin{equation}
\hspace{-10pt}
p({{x}'_{nt}}|\textbf{y}_t) \approx {\cal N}({{x}'_{nt}};{{\hat u}_{nt}},\nu _{nt}^u),
\end{equation}
\begin{equation}
\hspace{-49pt}
p({x_{nt}}|\sum\nolimits_r {{g_{nr}}{l_{rt}}} ) \approx {\cal N}({x_{nt}};{{\hat q}_{nt}},\nu _{nt}^q).
\vspace{-3pt}
\end{equation}
\noindent
Thus, the parameters ${\hat u_{nt}}(j),\nu _{nt}^u(j)$ computed during the final iteration of the M-GAMP phase, are treated as the prior parameters of $\textbf{X}$ in the L\&SPD phase. Conversely, the parameters ${\hat q_{nt}}(j')$ and $\nu _{nt}^q(j')$ computed during the final iteration of the L\&SPD phase are in turn used as the prior parameters of $\textbf{X}$ in the M-GAMP phase in (\ref{Xmarginal}).

In addition, ~to ~further ~exploiting ~the ~joint-sparsity ~of \vspace{-2pt}

\noindent
${\{x_{nt}\}}_{\forall n,t}$, we use the local support estimate ${{{\mathord{\buildrel{\lower3pt\hbox{$\scriptscriptstyle\leftharpoonup$}}\over \lambda } }_{nt}}}$ instead of the common sparsity rate $\lambda$ in (\ref{Xmarginal}). Then, by applying the SPA in the M-GAMP phase, we get
\vspace{-8pt}
\begin{equation}\label{lambdain}
\hspace{6pt}
\mathord{\buildrel{\lower3pt\hbox{$\scriptscriptstyle\leftharpoonup$}}
\over \lambda } _{nt}\!=\!\frac{{\lambda \prod\nolimits_{t' \ne t} {\mathord{\buildrel{\lower3pt\hbox{$\scriptscriptstyle\rightharpoonup$}}
\over \lambda } _{nt'}} }}{{\lambda \prod\nolimits_{t' \ne t} {\mathord{\buildrel{\lower3pt\hbox{$\scriptscriptstyle\rightharpoonup$}}
\over \lambda } _{nt'}}  + (1 - \lambda )\prod\nolimits_{t' \ne t} {(1 - \mathord{\buildrel{\lower3pt\hbox{$\scriptscriptstyle\rightharpoonup$}}
\over \lambda } _{nt'})} }},
\vspace{-4pt}
\end{equation}where the posterior local support probability
\vspace{-6pt}
\begin{equation}\label{lambdaout}
\hspace{-42pt}
{\mathord{\buildrel{\lower3pt\hbox{$\scriptscriptstyle\rightharpoonup$}}
\over \lambda } _{nt}} = {\left( {1 + \frac{{\mathcal{N}(0;{{\hat u}_{nt}},\nu _{nt}^u)}}{{\mathcal{N}({{\hat u}_{nt}};{{\hat q}_{nt}},\nu _{nt}^q + \nu _{nt}^u)}}} \right)^{ - 1}}.
\vspace{-6pt}
\end{equation}

\subsection{Algorithm Summary}
\vspace{-6pt}
\label{subsection3_3}
Beginning at the initial inter-phase iteration index, $i = 1$, the L\&S-AMP algorithm first performs the M-GAMP phase with the initial prior parameters ${\hat q_{nt}}=0,\nu_{nt}^q=1,{{{\mathord{\buildrel{\lower3pt\hbox{$\scriptscriptstyle\leftharpoonup$}}
\over \lambda } }_{nt}}}=0.5,~\forall n,t$ in (\ref{Xmarginal}). Then the converged outgoing messages $\{\hat u_{nt}\}_{\forall n,t},\{\nu _{nt}^u\}_{\forall n,t}$ are treated as prior parameters in the L\&SPD phase. Then the converged messages $\{\hat q_{nt}\}_{\forall n,t}$ and $\{\nu _{nt}^q\}_{\forall n,t}$ obtained from the L\&SPD phase, along with the updated beliefs $\{{{{\mathord{\buildrel{\lower3pt\hbox{$\scriptscriptstyle\leftharpoonup$}}
\over \lambda } }_{nt}}}\}_{\forall n,t}$ in (\ref{lambdain}), are used for the M-GAMP phase at inter-phase iteration $i = 2$. This procedure continues until either a stopping condition or a maximum number of allowable iterations is reached. Then we obtain the posterior mean estimates $\{\hat x_{nt}\}_{\forall n,t}$ computed in (\ref{Xestimate}).

Furthermore, we tune our prior and likelihood parameters $\left\{ {\lambda ,{{\hat g}_{0}},\nu _{0}^g, \{\tau _t^\omega\}_{\forall t}} \right\}$ using expectation-maximization \cite{AMPMMV13},\cite{EMGAMP11}, and estimate the rank $R$ using a rank selection strategy based on the penalized log-likelihood maximization in \cite{BiGAMP14}. In addition, we recommend initializing L\&S-AMP using $\lambda = 0.5$, ${{\hat g}_{0}} = 0, \nu _{0}^g = 1,$ and $\tau _t^\omega = 100, \forall t$.

\vspace{-10pt}
\section{Numerical Results}
\vspace{-8pt}
\label{section4}
In this section, we present real data results to compare the performance of the proposed L\&S-AMP algorithm with prior state-of-art PPXA \cite{L&Shypersensing12}, RA-ORMP \cite{RAORMP12}, SA-MUSIC \cite{SAMUSIC12}, and T-MSBL \cite{TMSBL11} algorithms. We evaluate the performance of the algorithms on two real hyperspectral datasets: 1) An urban dataset acquired over the University of Houston, with 144 spectral bands, $340\times740$ pixels, and a spatial resolution of $4m$. 2) An agricultural dataset acquired over the Salinas valley in California. The dataset has a spatial resolution of $3.7m$ and consists of $N=224$ spectral bands with each band corresponding to an image with $512\times217$ pixels.

We assume that the L\&S signal matrix $\textbf{X} \in {\mathbb{R}^{N \times T}}$ is obtained using pixel-based acquisition, so that $T$ denotes the number of pixels and $N$ denotes the number of spectral band. The DCT matrix is used as the sparsifying matrix $\bm{\Psi}$, and Gaussian noise is added to achieve SNR $= 25$ dB. It is worth noting that, for the sake of comparion, different random Gaussian measurement matrices $\textbf{A}_1, \textbf{A}_2,...,\textbf{A}_T$ are used. Also note that T-MSBL, RA-ORMP, and SA-MUSIC are derived only for the common measurement matrix case.

\begin{figure}[!t]
\setlength{\abovecaptionskip}{-1.5pt}
  \centering
  \includegraphics[width=8cm]{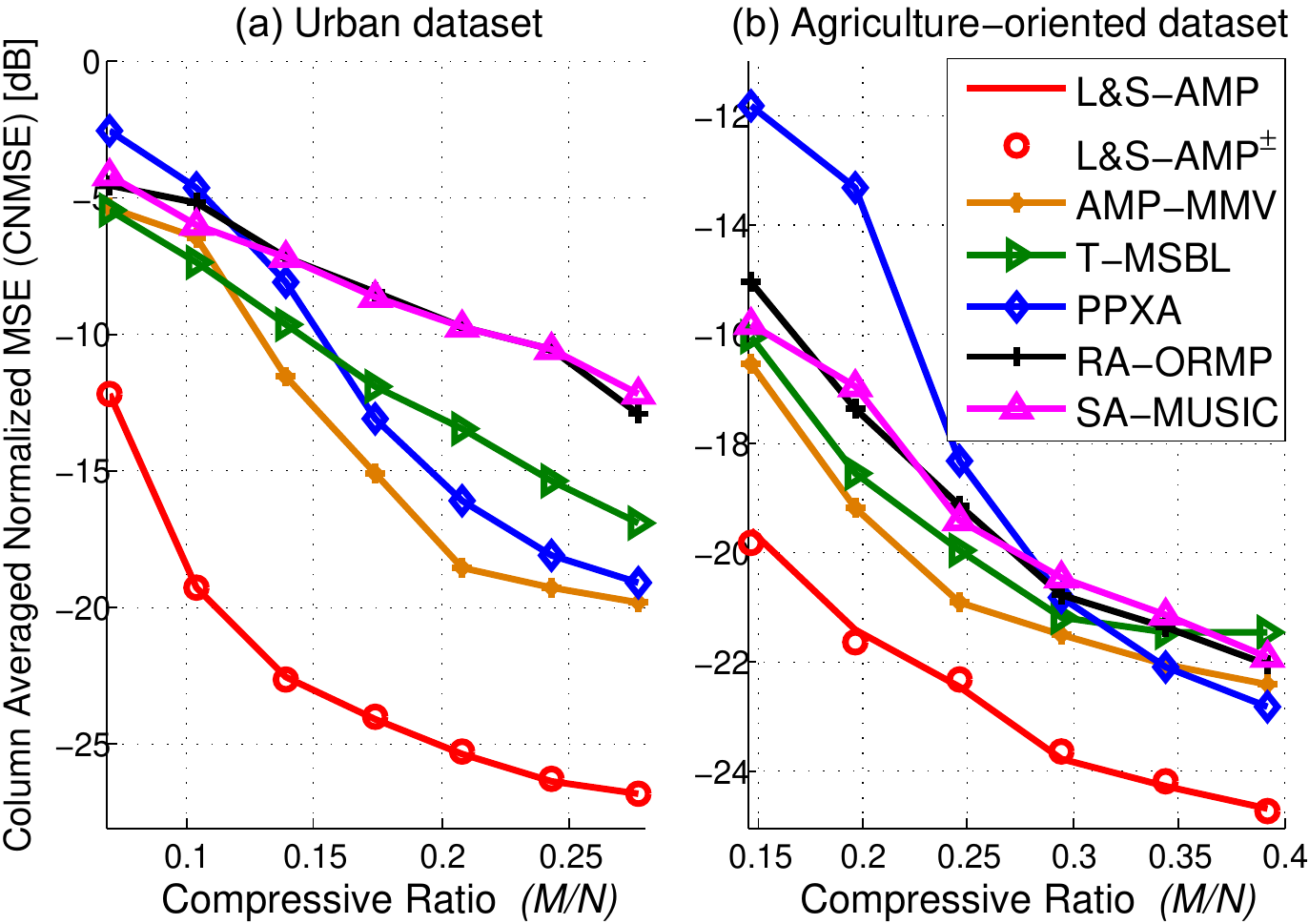}\\
\vspace{-10pt}
  \caption{CNMSE versus $M/N$ for the recovery of the urban dataset (left) and agriculture-oriented dataset (right).}\label{cnmse}
\vspace{-6pt}
\end{figure}

Fig. \ref{cnmse} plots the column-averaged normalized MSE (CNMSE) versus the compressive ratio $M/N$ on the two real datasets. The CNMSE is defined as ${\rm{CNMSE}}(\textbf{X},{{\textbf{\^ X}}}) = \frac{1}{T}\sum\nolimits_t {\left( {{{\left\| {{{\textbf{x}}_t} - {{\hat {{\textbf{x}}}}_t}} \right\|_2^2} \mathord{\left/{\vphantom {{\left\| {{{\textbf{x}}_t} - {{ {{\textbf{\^ x}}}}_t}} \right\|_2^2} {\left\| {{{\bf{x}}_t}} \right\|_2^2}}} \right.\kern-\nulldelimiterspace} {\left\| {{{\textbf{x}}_t}} \right\|_2^2}}} \right)}$, where ${\textbf{\^ X}}$ is an estimate of $\textbf{X}$. From the figure, we observe that the proposed algorithm outperforms all the other algorithms in terms of CNMSE, e.g., in Fig. \ref{cnmse}.(b), we note that L\&S-AMP achieves nearly 3dB reconstruction gain than the other algorithms at $M/N = 0.3$. In addition, a plus-minus sign ($\pm$) is used (i.e., L\&S-AMP$^\pm$) to denote the case of using random $\pm1$ measurement matrices, which are easy to implement in DMD, and can significantly reduce the burden of storage.

\begin{figure}[!t]
\setlength{\abovecaptionskip}{-1.5pt}
  \centering
  \includegraphics[width=8.4cm]{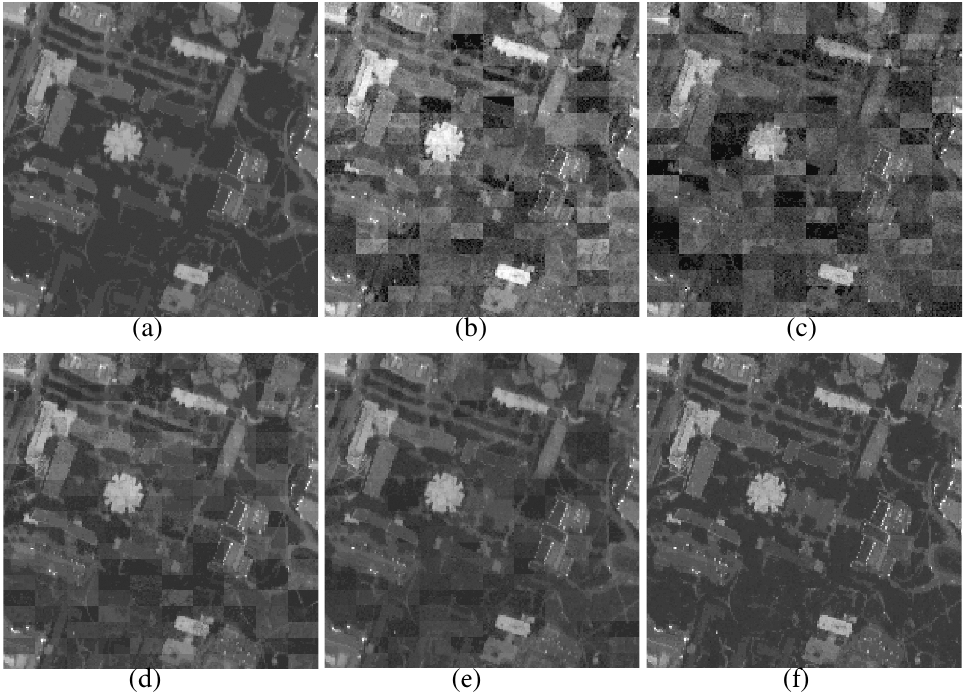}\\
\vspace{-12pt}
  \caption{Visual quality comparison of the recovered images for the urban dataset. From left to right and top to bottom: original image, the recovered images by SA-MUSIC, RA-ORMP, T-MSBL, PPXA, and the proposed algorithm. $M/N$ is fixed to 0.243, and other simulation parameters remain unchanged. The whole scene is partitioned into a sequence of sub-scenes to enable parallel processing. }\label{visual}
\vspace{-10pt}
\end{figure}

Some visual results of the recovered hyperspectral images by using different algorithms are presented in Fig. \ref{visual}. As expected, our proposed algorithm preserves more fine details and much sharper edges, and shows much clearer and better visual results than the other competing methods.

\vspace{-10pt}
\section{Conclusion}
\vspace{-8pt}
\label{section5}
In this paper, we studied joint CS reconstruction of spatially and spectrally correlated hyperspectral data acquired, assuming that the hyperspectral signal matrix satisfies the joint-sparse model with a lower rank than the sparsity level, i.e., the L\&S model. We proposed an AMP-based algorithm for recovering the signal matrix with the L\&S model while exploiting the structured sparsity and the amplitude correlation of the data. The numerical results were presented to confirm the performance advantage of our algorithm.

\end{document}